\documentclass[12pt, prd, showpacs]{revtex4}
\usepackage{amsmath}
\usepackage{amssymb}

\begin{document}

\title{Zel'dovich states with very small mass and charge in nonlinear
electrodynamics coupled to gravity}
\author{O. B. Zaslavskii}
\affiliation{Astronomical Institute of Kharkov V.N. Karazin National University, 35
Sumskaya St., Kharkov, 61022, Ukraine}
\email{zaslav@ukr.net}

\begin{abstract}
It is shown that in non-linear electrodynamics (in particular, Born-Infeld
one) in the framework of general relativity there exist "weakly singular"
configurations such that (i) the proper mass $M$ is finite in spite of
divergences of the energy density, (ii) the electric charge $q$ and
Schwarzschild mass $m\sim q$ can be made as small as one likes, (iv) all
field and energy distributions are concentrated in the core region. This
region has an almost zero surface area but a finite longitudinal size $L=2M$%
. \ Such configurations can be viewed as a new version of a classical
analogue of an elementary particle.
\end{abstract}

\maketitle




\section{Introduction}

There is a long-standing problem of Coulomb divergences in classical
electrodynamics. In a flat space-time, this drawback can be remedied in
non-linear electrodynamics where corresponding localized soliton-like
solutions replace point-like solutions of Maxwell theory \cite{bi}. These
solutions represent extended objects which are free of divergences and can
be considered as candidates to the role of classical analogues of an
elementary particles. Meanwhile, in general relativity the situation
changes. There exists a general theorem \cite{bs1}, \cite{bs2}, \cite{bcom}, 
\cite{bkd} that states that a regular center is impossible, provided the
nonlinear Lagrangian behaves like the Maxwell one in the limit of weak
fields. This difficulty can be avoided by making Lagrangians inhomogeneous
with non-Maxwellian behavior near the centre \cite{plb} that looks, however,
artificial. Although the nonzero global electric charge forbids the
existence of globally regular solutions \cite{bkd}, some properties of
electrically charged configurations deserve study.

First, it turns out that although globally regular solutions are impossible,
one can sometimes achieve the finiteness of the proper mass in spite of the
singular centre that represent so-called "weakly singular" solutions (see,
e.g., textbook \cite{sh}). Second, it turns out that among such solutions
there exists a particular class with rather unusual interesting properties:
their Schwarzschild mass $m$ almost vanishes, the electric charge $q$ so
does but the proper mass $M$ is still finite and non-zero. In this limit,
all non-trivial field configuration is localized near the centre under the
sphere of an almost vanishing radius but with finite longitudinal size. As a
result, we obtain a classical analogue of an elementary particle which is in
a sense combines features of a point-like and extended objects. For an
external observer it reveals itself as almost "nothing" and, in this
respect, is similar to a friedmon \cite{fried}. However, in contrast to it,
there is no "other Universe" inside, all game is developed in the same space
which is static everywhere.

In the present paper we describe the solutions with aforementioned
properties. We do not pretend, of course, for comparison of the parameters
of the obtained configurations with experimental data. Our goal is rather
methodical: to demonstrate some unusual features that general relativity
brings into non-linear electrodynamics and which have no analogues in flat
space-times. Nonetheless, we would like to remind that, since, say, for an
electron $m\ll q$ in geometrical units, this corresponds to a naked
singularity that can be considered at least as an additional motivation for
studying distributions which are singular (or smooth but become singular in
some limits). The example of studies of such a kind in which a singular
configuration was considered as a model of elementary particle within
general relativity and linear electrodynamics, can be found in \cite{bon}.

The configurations discussed in our paper are obtained on the basis of the
energy distribution considered by Zel'dovich a long time ago \cite{z} in a
quite different context connected with relativistic astrophysics (the
collapse of relativistic star of a small mass in general relativity). The
similar distributions were also discussed in the context of black \ hole
thermodynamics \cite{ber-th}.

\section{Zel'dovich's configurations and ultimate gravitation mass defect}

Let us consider the spherically-symmetric metric%
\begin{equation}
ds^{2}=-dt^{2}\exp (2\gamma )+dr^{2}\exp (2\alpha )+\exp (2\beta )d\omega
^{2}\text{, }d\omega ^{2}=d\theta ^{2}+\sin ^{2}\theta d\phi ^{2}\text{.}
\end{equation}%
It follows from the $00$ component of the Einstein equations that%
\begin{equation}
\exp (-2\alpha )=1-\frac{2m(r)}{r}\text{,}
\end{equation}%
\begin{equation}
m(r)=4\pi \int_{0}^{r}d\bar{r}\bar{r}^{2}\rho \text{, }m(\infty )\equiv m,
\label{m}
\end{equation}%
where $\rho $ is the energy density$,m$ has the meaning of the Schwarzschild
mass. We assume that there is no horizon, so that $r>2m(r)$, $\exp (-2\alpha
)>0$ for all $r>0$.

The proper mass%
\begin{equation}
M=4\pi \int_{0}^{\infty }dr\exp (\alpha )\rho r^{2}\text{.}  \label{mp}
\end{equation}

Now let us consider the situation in which for $r\leq r_{0}$ the energy
density has the form \cite{z}%
\begin{equation}
\rho =\frac{b}{8\pi r^{2}}\text{,\thinspace }  \label{rod}
\end{equation}%
where $b$ is a numerical coefficient, and for $r>r_{0}$ $\rho =0$. Then, for 
$r\leq r_{0}$ 
\begin{equation}
\exp (-2\alpha )=1-b
\end{equation}%
where we assume $b<1$ to ensure $\exp (-2\alpha )>0$. For $r\leq r_{0}$, the
mass function reads 
\begin{equation}
m(r)=\frac{br}{2}\text{, }m=\frac{br_{0}}{2}\text{.}  \label{adm}
\end{equation}%
In doing so, the proper mass 
\begin{equation}
M=\frac{br_{0}}{2\sqrt{1-b}}  \label{M}
\end{equation}%
It can be rewritten as%
\begin{equation}
M=\frac{bL}{2}\text{, }L=\frac{r_{0}}{\sqrt{1-b}}  \label{dl}
\end{equation}%
where $L$ is the proper length that characterizes the longitudinal size of
the system from the center to $r_{0}$. It is also worth noting a simple
bound that follows from (\ref{dl}):%
\begin{equation}
M\leq \frac{L}{2}\text{.}  \label{bound}
\end{equation}

In general, $r_{0}$ and $b$ are two independent parameters. However, if one
arranges the special relationship between them, it is possible to obtain
non-trivial realization of the gravitational mass defect. This was shown by
Zel'dovich \cite{z} for the ultrarelativistic fermi gas of nucleons with the
equation of state 
\begin{equation}
\rho \sim n^{4/3}
\end{equation}%
where $n$ is the number density of nucleons. Their total number is equal to%
\begin{equation}
N=4\pi \int drr^{2}\exp (-\alpha )n\text{.}  \label{N}
\end{equation}

Then, it follows from (\ref{m}), (\ref{mp}) (\ref{rod}) and (\ref{N}) that%
\begin{equation}
r_{0}\sim N^{2/3}(1-b)^{1/3}\text{, }m\sim N^{2/3}(1-b)^{1/3}\text{, }M\sim
N^{2/3}(1-b)^{-1/6}\text{.}  \label{prop}
\end{equation}

Thus, taking the baryon number $N$ to be fixed we obtain that for $1-b\ll 1$
the quantities $r_{0}$and $m\approx \frac{r_{0}}{2}$ can be made as small as
one wishes.$.$

Zel'dovich's goal was to show that one can always rearrange the distribution
of a fixed number of nulceons to achieve an almost vanishing gravitational
mass $m$. Meanwhile, it remained unnoticed in \cite{z} that for such a
configuration the proper mass $M\rightarrow \infty $. This is interesting by
itself since supplies us with the manifestation of the ultimate
gravitational mass defect: the Schwarzschild mass vanishes but the proper
mass diverges. Earlier, another example of such a situation was reported in 
\cite{ult} but it relied heavily on non-stationary geometries. Meanwhile,
now the space-time is pure static, so one obtains one more type of the
ultimate gravitational mass defect.

\section{Modification of Zel'dovich configuration}

Being interesting for astrophysical applications \cite{li}, the
configurations suggested in \cite{z} are not suitable for our purposes to
find an analogue of classical elementary particle in nonlinear
electrodynamics. The proper mass for configurations considered in \cite{z}
diverges as we saw above. In principle, this can be remedied if, instead of $%
N$ (which does not have sense at all for the problem under discussion), we
just keep $M$ fixed by hand. However, this is insufficient since the density
has somewhat different profile in all relevant electrodynamic theories (see
next section). This leads to modification of the original Zel'dovich's
distribution. Namely, let we have two typical regions: region I where $%
r\rightarrow 0$ (more exactly, $r<r_{0}$ where $r_{0}$ characterizes the
scale on which density changes, $r_{0}\rightarrow 0$ in final formulas) and
region II where $r\gtrsim r_{0}$. In region I we suppose the validity of the
asymptotic form of the type (\ref{rod}). However, we take into account in $%
\rho $ not only the divergent term but the first small correction in
expansion with respect to inverse powers of $r$. More specifically, we
suppose that there are no terms of $r^{-1}$ (as it happens in physically
relevant applications discussed below), so instead of (\ref{rod}) now 
\begin{equation}
\rho \approx \rho _{1}-\rho _{2}\text{, }\rho _{1}=\frac{b}{8\pi r^{2}}\text{%
, }\rho _{2}=\frac{1}{8\pi \lambda }  \label{roa}
\end{equation}%
where $\lambda >0$ is a constant. The role of $r_{0}$ is played by the
parameter $\lambda $, $\rho _{2}\ll \rho _{1}$. We will see below that such
a dependence arises, in particular, in nonlinear Born-Infeld
electrodynamics. In region II we suppose that $\rho =\varepsilon f(r)$ where
the parameter $\varepsilon \rightarrow 0$ and $f$ is the bounded function.
In particular, the role of $\varepsilon $ can be played by an electric
charge. Then, for $\varepsilon \rightarrow 0$ it is region I which gives the
main contribution to the mass.

Then, for $r$ $\rightarrow 0$ 
\begin{equation}
\exp (-2\alpha )\approx 1-b+\frac{r^{2}}{3\lambda }\text{.}
\end{equation}%
Now we want to examine under what conditions the proper energy $M$ converges
in the situation when $r_{0}$ and $m$ can be made arbitrarily small. We
assume that at infinity $\rho $ falls off rapidly enough, so possible
divergences are connected with the lower limit $r=0$ of the integral (\ref%
{mp}) only. If $b\neq 1$, the integral trivially converges. However, when
formally $b\rightarrow 1$, in general $M\rightarrow \infty $. In doing so,
the contribution from the constant term in (\ref{roa}) is finite and
proportional to $r_{0}$. Let us evaluate the dominant contribution $M_{1}$
which stems from the first term in (\ref{roa}). It behaves like%
\begin{equation}
M_{1}\approx \frac{\sqrt{3\lambda }}{2}[D+\frac{1}{2}\ln \frac{1}{\lambda
(1-b)}]  \label{md}
\end{equation}%
where the exact value of the constant $D$ is irrelevant for our purposes. If 
$\lambda $ remains finite, $M_{1}\rightarrow \infty $ when $b$ approaches
the value $b=1$. However, let us try to adjust two transitions $\lambda
\rightarrow 0$ and $b\rightarrow 1$ in (\ref{md}) in such a way that the
quantity $M_{1}$ remain finite. It is indeed possible if 
\begin{equation}
1-b\approx \frac{B}{\lambda }\exp (-\frac{4a}{\sqrt{3\lambda }})  \label{1d}
\end{equation}%
where $B$ and $a$ are constants. In other words, if $\lambda \rightarrow 0$
and $b\rightarrow 1$ along the curve (\ref{1d}) with an arbitrary value of
constant $B$, the mass $M$ remains finite. More precisely, it follows from
the substitution of (\ref{1d}) back into (\ref{md}) that $M_{1}=a+O(\sqrt{%
\lambda })$. The choice of the constant $B$ results in the corrections in $%
M_{1}$ absorbed by $D$ as it is clear from (\ref{md}). However, in the limit 
$\lambda \rightarrow 0$ any such corrections are multiplied by the small
factor proportional to $\sqrt{\lambda }$, so in this limit it does not
affect the value of $M_{1}\rightarrow a$. To be accurate, we must check that
the quantity $M_{2}$ does not generate new divergences in this limit.
Indeed, the core region I gives the contribution $M_{2}\sim \frac{r^{2}}{%
\sqrt{\lambda }}\sim \sqrt{\lambda }\frac{r^{2}}{\lambda }$ represents a
product of two small factors and is, obviously, finite and arbitrarily
small. As a result, we obtain the configuration for which $M\rightarrow
const\neq 0$,$\frac{m}{M}$ becomes arbitrarily small. Now we will see that
such a situation can arise in nonlinear electrodynamics.

For all configurations under discussion the proper distance $L$ behaves in
the way similar to $M$ since according to (\ref{mp}) both quantities differ
from each other by the factor $\rho r^{2}\,\ $which is finite according to (%
\ref{rod}). Moreover, for $r_{0}\ll M$, $1-b\ll 1$ we obtain the relation $M=%
\frac{L}{2}$ that agrees with (\ref{dl}).

\section{Nonlinear electrodynamics: basic formulas}

In this section, we consider general basic formulas of nonlinear
electrodynamics needed for what follows. In doing so, we follow the
presentation of \cite{sh}. Consider the self-gravitating system with
Lagrangian

\begin{equation}
L=-\frac{R}{16\pi }-\Phi (I)\text{,}
\end{equation}%
where $R$ is the Riemann curvature and we use the units with $G=c=1$. Here,
the field invariant%
\begin{equation}
I=-F_{\alpha \beta }F^{\alpha \beta }
\end{equation}%
where $F_{\alpha \beta }$ is the tensor of the electromagnetic field. Then,
the stress-energy tensor takes the form%
\begin{equation}
T_{\mu }^{\nu }=diag(-\rho ,p_{r},p_{\perp },p_{\perp })\text{,}
\end{equation}%
where%
\begin{equation}
\rho =-p_{r}=2I\Phi _{I}-\Phi \text{, }\Phi _{I}=\frac{d\Phi }{dI}\text{, }%
p_{\perp }=\Phi  \label{pr}
\end{equation}%
It follows from $00$ and $11$ Einstein equations that (up to the additive
constant) that for our system%
\begin{equation}
\gamma =-\alpha \text{.}  \label{01}
\end{equation}

The Maxwell equations read%
\begin{equation}
\frac{1}{\sqrt{-g}}\frac{\partial (\sqrt{-g}F^{\alpha \beta }\Phi _{I})}{%
\partial x^{\beta }}=0\text{.}  \label{max}
\end{equation}%
We assume that the system is static and spherically-symmetric. Then, the $0$%
-component of (\ref{max}) together with (\ref{01}) gives us that 
\begin{equation}
\Phi _{I}F^{0r}=\frac{q}{16\pi r^{2}}  \label{fi}
\end{equation}%
where $q$ is an electric charge (for definiteness, in what follows we choose 
$q>0$). Now, $I=-2F_{0r}F^{0r}=2\left( F^{0r}\right) ^{2}>0.$ With our sign
convention, $\Phi >0$. By the integration of (\ref{fi}) we obtain that%
\begin{equation}
\Phi =\frac{\sqrt{2}q}{16\pi }\int_{0}^{I}\frac{dI}{r^{2}\sqrt{I}}
\end{equation}%
where we took into account that at $r\rightarrow \infty $ $I\rightarrow 0$.
Integrating by parts and taking into account (\ref{pr}), one obtains that%
\begin{equation}
\rho =\frac{q\sqrt{2}}{4\pi }\int_{r}^{\infty }\frac{d\bar{r}\sqrt{I}}{\bar{r%
}^{3}}\text{.}
\end{equation}

Thus, if $I\leq \infty $ in the centre where $r\rightarrow 0$, the energy
density has the universal behavior $\rho \approx \frac{A}{r^{2}}+const$
where $A\sim q.$ This is just the behavior typical of the Zel'dovich
gravitational configurations \cite{z} or their generalization (\ref{roa}).
However, in general, for $q\neq 0$, the Schwarzschild mass remains nonzero.
Only in the situation when $q$ becomes arbitrarily small and some other
parameters are fine-tuned (see details below) one does indeed obtain the
configuration with arbitrarily small $m$ and finite $M$.

\section{Born-Infeld electrodynamics}

Let the system have the electrodynamic Lagrangian

\begin{equation}
\Phi _{BI}=\frac{1}{8\pi \lambda }(1-\sqrt{1-\lambda I})\text{.}
\end{equation}%
Then, applying formulas of the previous section, we obtain%
\begin{equation}
I=\frac{2q^{2}}{r^{4}+2q^{2}\lambda }\text{,}
\end{equation}%
\begin{equation}
\rho =\frac{1}{8\pi \lambda r^{2}}(\sqrt{r^{4}+2\pi \lambda q^{2}}-r^{2})%
\text{.}
\end{equation}%
The value of the Schwarzschild mass (\ref{m}) 
\begin{equation}
m\approx 0.84q\text{.}
\end{equation}%
Thus, for $r\gg \left( 2\pi \lambda q^{2}\right) ^{1/4}\equiv r_{0}$ we have
the standard asymptotics%
\begin{equation}
\rho \approx \frac{q^{2}}{8\pi r^{4}}  \label{rq}
\end{equation}%
typical of the linear electrodynamics whereas for $r\ll r_{0}$ the density
has the form (\ref{roa}) with%
\begin{equation}
b=\sqrt{\frac{\lambda _{\min }}{\lambda }}\text{, }\lambda _{\min }\equiv q%
\sqrt{2}\text{.}
\end{equation}

If $\lambda \rightarrow \lambda _{\min }$, $r_{0}\sim q\sim \lambda $ in
accordance with general consideration in Sec. III. In general, in this limit
the proper energy $M\rightarrow \infty $. However, according to (\ref{1d}),
it remains finite provided $q\rightarrow 0$ and the parameters are
fine-tuned according to%
\begin{equation}
\sqrt{\lambda }-2^{1/4}\sqrt{q}\approx \frac{2^{-1/4}B}{3\sqrt{q}}\exp (-%
\frac{2^{7/4}M_{1}}{\sqrt{3q}})  \label{1}
\end{equation}%
$M_{1}$ and $B$ being finite fixed quantities.

It is worth noting that in this limit, for any intermediate value of $r$ the
density has the form (\ref{rq}) and, thus, tends to zero uniformly. In the
core region $r<r_{0}$ it diverges but the size of the core $r_{0}$ shrinks
in this limit. As a result, the Schwarzschild mass $m$ becomes arbitrarily
small but $M$ remains finite nonzero quantity. This is true in spite of the
fact that the invariant $I(0)\sim q^{-1}$ diverges in this limit as well as
the invariant $\Phi (0)$.

\section{Nonlinear Schr\"{o}dinger Lagrangian}

Let now

\begin{equation}
\Phi _{Sch}=-\frac{1}{8\pi \lambda }\ln (1-\frac{\lambda I}{2}),
\end{equation}%
\begin{equation}
\text{ }\rho =\frac{1}{8\pi \lambda r^{2}}[\frac{4\lambda q^{2}}{f}+r^{2}\ln
(\frac{2r^{2}}{f})]\text{, }I=\frac{8q^{2}}{f^{2}}\text{, }f=\sqrt{%
r^{4}+4\lambda q^{2}}+r^{2}\text{.}
\end{equation}

Thus, at $r\rightarrow 0$ the energy density does indeed has the asymptotic
form (\ref{roa}) with 
\begin{equation}
b=\frac{2q}{\sqrt{\lambda }}\equiv \sqrt{\frac{\lambda _{\min }}{\lambda }}%
\text{.}
\end{equation}%
This example has some specific additional features as compared to the
previous ones because of the logarithmic term that needs the modification of
the described scheme but this difference is not crucial. Then, 
\begin{equation}
\exp (2\gamma )\approx 1-b-r^{2}\frac{\ln \frac{r^{2}}{\sqrt{\lambda q^{2}}}%
}{3\lambda }\text{, }m\approx 0.8q\text{.}
\end{equation}
The potentially dangerous term in the proper mass has the form (we omit for
simplicity numerical coefficients and give rough estimate)%
\begin{equation}
M\sim \sqrt{\lambda }\left\vert \ln \left\vert \ln x\right\vert \right\vert 
\text{, }x=\frac{1-b}{\left\vert \ln (1-b)\right\vert }
\end{equation}

Thus, the finiteness of $M$ is possible provided the parameters are adjusted
in such a way that%
\begin{equation}
\frac{1-b}{\left\vert \ln (1-b)\right\vert }\sim \exp [-\exp (\frac{M}{\sqrt{%
\lambda }})]\text{.}
\end{equation}

\section{Discussion and conclusion}

The configurations under discussion represent counterpart of those in \cite%
{z} for a system with long-range forces. They share a rather unusual
interesting property. Namely, when parameters of the system are fine-tuned
in certain way, both the Schwarzschild mass $m$ and charge $q\sim m$
approach zero as closely as one likes but the proper mass $M$ remains
nonzero. It is worth stressing that one cannot simply put $m=0$ or $q=0$ (or
both) from the very beginning - this would have completely destroyed such
configurations. In doing so, the imprint on space-time from the electric and
gravitation fields disappears outside. All remaining fields are concentrated
in the core region. In doing so, the core region has an almost zero surface
area but with a longitudinal size $L=2M$, so that a finite string-like
object is obtained. Thus, although our particle-like configuration is
localized within a sphere with an arbitrarily small size, it cannot be made
point-like in agreement with discussion of other classical analogs of an
elementary particle \cite{mf}.

The configurations in question realize the strong gravitational mass defect
since the ratio $\frac{m}{M}$ can be made as small as one wishes. It is also
worth noting that, when the charge diminishes, the value of the field
invariant in the centre $I(0)\sim q^{-1}$ grows unbound.

To some extent, the configurations under discussion resemble a semi-closed
world \cite{fried} but, instead of the whole Universe inside a "particle",
now we have a weak singularity in the same space which almost disjoints from
the outer world.

Our consideration was pure classical. As is known, account for quantum
effects places severe limitations on the relevance of Zel'dovich's
configurations in relativistic astrophysics \cite{mar}, \cite{ber}. It would
be of interest to carry out similar analysis for the systems considered in
the present work. It is also of interest to extend the approach under
discussion to the non-spherical (say, axially-symmetric) case when
non-trivial static solutions with $m\rightarrow 0$ are known to exist \cite%
{f}. A separate issue is the analysis of rotating configurations.

Also, other approaches to the problem of singularities of a point-like
charge including quantum field theory effects can be of interest (see, e.g.,
the recent work \cite{vd})

I thank K.\ A. Bronnikov for useful comments.

\end{document}